\documentclass[prl,reprint,twocolumn,letterpaper,groupedaddress,amsmath,amssymb,showpacs,aps,english,superscriptaddress]{revtex4-1}

\usepackage{subfigure}
\usepackage{graphicx}
\usepackage{bm,mathrsfs}
\usepackage{amsmath}
\usepackage{amssymb}
\usepackage{esint}
\usepackage{setspace}
\usepackage{hyperref}
\usepackage{color,colortbl}
\usepackage{colortbl}
\usepackage{mathtools}
\usepackage{multirow}
\usepackage{bbold}
\usepackage{pbox}
\usepackage{todonotes}
\usepackage{hhline}
\usepackage{comment}
\usepackage{tabularx}
\usepackage{cancel}
\usepackage{placeins}
\allowdisplaybreaks
\newcolumntype{C}[1]{>{\centering\arraybackslash}p{#1}}

\makeatletter

\begin{document}

\title{Quantum metrology with a single spin-$3/2$ defect in silicon carbide}

\author{\"O. O. Soykal}
\email[Email: ]{oneysoykal@gmail.com}
\author{T. L. Reinecke}
\affiliation{Naval Research Laboratory, Washington, DC 20375}
\begin{abstract}
We show that implementations for quantum sensing with exceptional sensitivity and spatial resolution can be made using spin-3/2 semiconductor defect states. We illustrate this using the silicon monovacancy deep center in hexagonal SiC based on our rigorous derivation of this defect's ground state and of its electronic and optical properties. For a single $\textrm{V}_{\textrm{Si}}^-$ defect, we obtain magnetic field sensitivities capable of detecting individual nuclear magnetic moments. We also show that its zero-field splitting has an exceptional strain and temperature sensitivity within the technologically desirable near-infrared window of biological systems. The concepts and sensing schemes developed here are applicable to other point defects with half spin multiplet ($S\geq 3/2$) configuration.
\end{abstract}

\maketitle

Technologies based on quantum information are recently opening a range of new opportunities from secure communications to quantum computing. Quantum sensing using entangled entities such as spins, atomic excitations, and photons can provide vastly improved sensitivities compared to classical technologies. Sensing using defect spin states in semiconductors is particularly important in part because of its potential for high spatial resolution and for integration with existing solid-state technologies \cite{Weber_PNAS10,Togan_nature11,Bernien_Nature13,Grinolds_NP13,Balasubramanian_Nature08,Shi_science15}. Room temperature magnetic and strain sensing are being currently investigated using spin-1 and inter-valley spin states, e.g. nitrogen-vacancy (NV) deep color centers in diamond \cite{Dolde_NP11} and phosphorous shallow donors in silicon \cite{Lo_nmat15,Soykal_PRL11}, that require difficult micro-fabrication processes and experimentally challenging detection techniques.

New concepts and approaches have the potential to move quantum sensing forward to higher sensitivities in systems that are easier to implement. In the present work we show that defect states with less common spin-$3/2$ (or other half spin multiplets) ground state configuration provide qualitatively a unique opportunity in quantum sensing due to unusual entanglement properties of their spin states, reduced losses, and Kramer’s degeneracy. To achieve this, we address the spin-$3/2$ $\textrm{V}_{\textrm{Si}}^-$ monovacancy center \cite{janzen_physicab09,Kraus_NP14,Widmann_nmat15,Carter_PRBRC15,Soykal_PRB2016} in hexagonal SiC and develop novel sensing schemes resulting in extraordinary sensitivities in magnetic, strain, and temperature sensing. We note that the technologically important wide band gap silicon carbide (SiC) \cite{Koehl_nature11, Falk_natcom13,Christle_nature14,Son_PRL06,Falk_PRL15} has mature growth and microfabrication technologies and favorable optical emission wavelengths \cite{Song_OptExp11,Maboudian_JVSci13,Baranov_prb11}, and we develop optical sensing protocols that are particularly easy to implement.

For the $\textrm{V}_{\textrm{Si}}^-$ defect, we find an unexpected avoided crossing of its GS spin states forming a naturally entangled $\Lambda$-type system leading to a significant increase in sensitivity to magnetic fields. Such an avoided crossing has been observed recently \cite{Simin_arxiv}. The degeneracy in these entangled spin states allows for coherent control by using static magnetic fields. In addition, we obtain an important relationship between its GS zero-field splitting (ZFS) and strain coupling that can be employed for on-chip strain detection using realistic SiC micro-electro-mechanical systems (MEMS). We also show that its GS-ZFS is highly sensitive to temperature and can be used for bio-chemical sensing either optically in the desirable near-infrared window of biological systems or paramagnetically with current magnetic resonance imaging technology.

\begin{figure}[!ht]
\centering
\includegraphics*[width=7. cm]{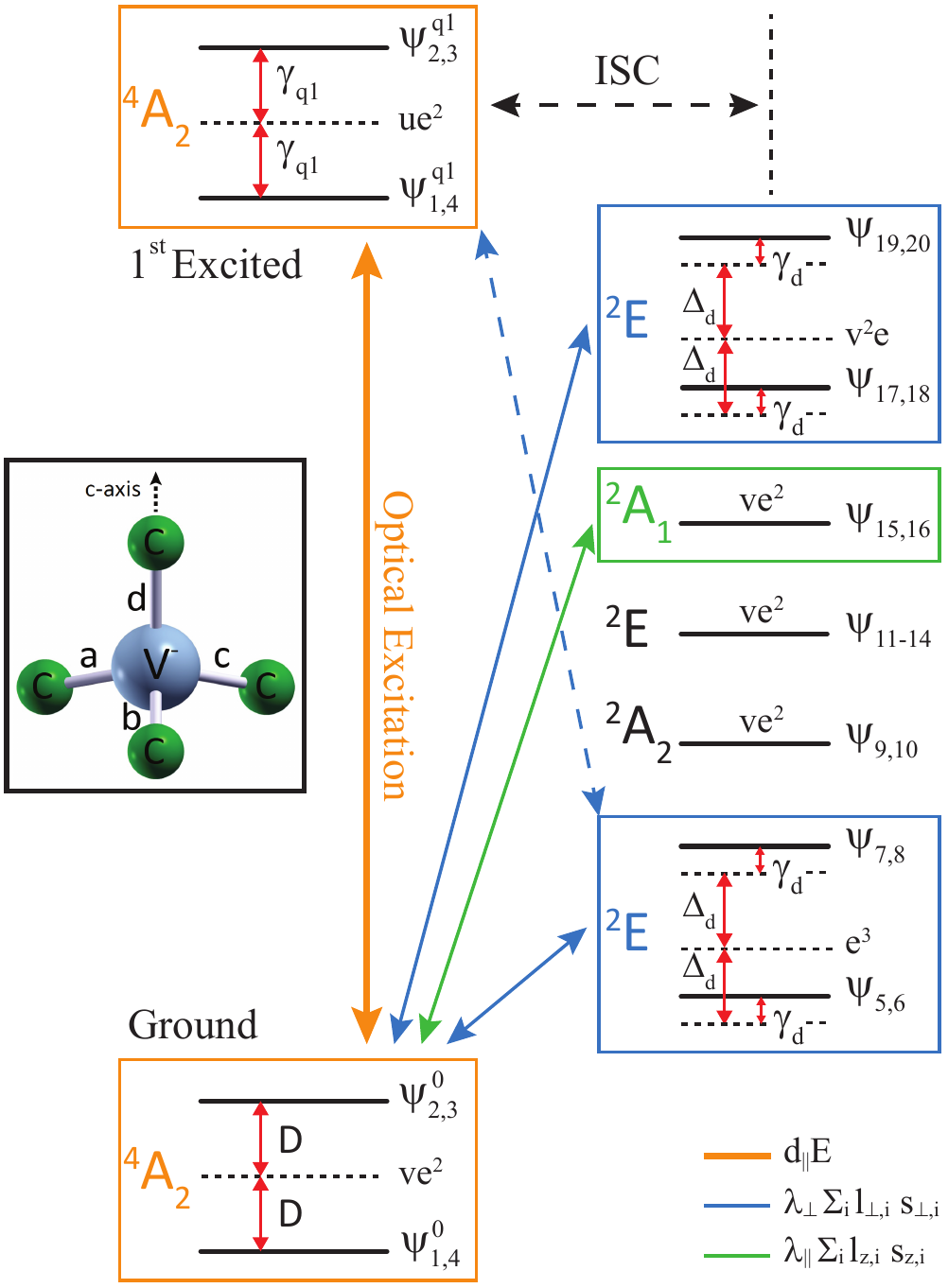}
\caption{Electronic structure and wave functions of $V_{\textrm{Si}}^-$ in 4H-SiC. Each state has a 2-fold Kramer's degeneracy. The quartet GS ($\prescript{4}{}A_2$) can be optically spin polarized and read-out via the first quartet ES ($\prescript{4}{}A_2$) with $d_{||}$ dipole moments along the defect's c-axis. The dark doublet states (3 $\prescript{2}{}E$, $\prescript{2}{}A_1$, and $\prescript{2}{}A_2$) are ordered in energy on the right. Non-vanishing spin-orbit matrix elements to GS (ES) are shown by solid (dashed) arrows forming an ISC channel between quartets and doublets. $\Delta_d$ and $\gamma_d$ are the energy splittings/shifts induced by the SO and SS interactions. ZFS splittings are labeled $2D$ for the GS and $2\gamma_{q1}$ for the ES. (Inset) Local $C_{3\nu}$ symmetry of the defect. $a,b,c,d$ represent the $sp^3$ dangling bonds of the surrounding carbon atoms.}\label{Fig1}
\end{figure}

The silicon \textit{monovacancy} $\textrm{V}_{\textrm{Si}}^-$ in hexagonal silicon carbide (4H-SiC) is a point defect with $C_{3\nu}$ symmetry consisting of a negatively charged silicon vacancy surrounded by four carbon (C) atoms (see inset of Fig.\ref{Fig1}). It has five active electrons, four from the $sp^3$ dangling bonds of C atoms and one from the extra charge. Its electronic structure up to the first optically active excited state (ES) is shown in Fig.\ref{Fig1}. Its GS has a quartet ($S=3/2$) spin configuration with a zero field splitting of $2D \sim 70$ MHz \cite{Carter_PRBRC15} between the spin $m_s=\pm 3/2$ (lower) and $m_s\pm 1/2$ (higher) states due to the spin-spin interactions \cite{Soykal_PRB2016}. Optical excitation from GS to ES, both with $\prescript{4}{}A_2$ symmetry, is allowed for an electric dipole moment parallel to the c-axis of the defect. The dark doublet states are coupled to the GS and ES quartets through the spin-orbit interactions giving a spin-selective radiationless decay path --known as the inter-system crossing (ISC). Through this ISC, the ES can transition radiationlessly back to GS with different rates for each spin multiplicity $m_s{=}\pm3/2$ and $m_s{=}\pm 1/2$. This leads to the spin polarization of the GS. After a steady-state is reached, spin-dependent changes in the photoluminescence (PL) will occur when populations are modified.

A rigorous, fully relativistic, multi-particle derivation of its ground state (GS) spin Hamiltonian, including the spin-orbit (SO) and spin-spin (SS) interactions, is needed here as a basis for novel sensing protocols. To obtain such a Hamiltonian, we apply perturbation theory to the GS wave functions using the SO potential \cite{Stoneham1985} $V_{so}{=}\sum_{i} \lambda_{||} l_{z,i}s_{z,i}{+}\lambda_{\bot}(l_{x,i}s_{x,i}+l_{y,i}s_{y,i})$. Orthogonal and longitudinal SO coupling parameters along the basal plane and c-axis are $\lambda_\bot$ and $\lambda_{||}$, respectively. Using symmetry-adapted multi-particle wave functions \cite{Soykal_PRB2016} expressed in terms the molecular orbitals (MOs), we obtain the SO corrected ground state wave functions (see Fig.\ref{Fig1}), $\Psi_i^{so}{=}\Psi_i^0{+}\sum_j\alpha_{i,j}\Psi_j$, up to the first order perturbation coefficients $\alpha_{i,j}$ \cite{supplement}. The interaction between the GS spins and a magnetic field is given by the fully relativistic multi-particle Hamiltonian \cite{Stone1963} $H_{B}=\sum_i \mu_B(\bm{l}_i+g_e\bm{s}_i)\bm{B}/\hbar+\sum_i e^2\left(\bm{B}\times\bm{d}_i\right)^2/8mc^2+e\sum_{i,j}\left(\bm{s}_{i}\times\nabla_r V_k(\bm{r}_{ij})\right)\left(\bm{B}\times\bm{d}_i\right)/4m^2 c^3$, where $i$ and $j$ are electron and nuclear indices. $g_e$, $\mu_B$, $e$, $m$, and $c$ are the bare electron g-factor, Bohr magneton, electron charge and mass, and speed of light, respectively. The $i^\textrm{th}$ electron's position relative to the $j^\textrm{th}$ nucleus is given by $\bm{r}_{ij}$. The position vector of the electron relative to an arbitrary origin is $\bm{d}$, and $\bm{l}{=}\bm{d}{\times}\bm{p}$ is the angular momentum about this origin. Thus, the first term corresponds to the Lande g-factor. The second term, proportional to $\bm{B}^2$ and independent of the spin, shifts the energy levels, and it can be omitted. The last term is the relativistic correction to the first term due to the nuclear potentials $V_k$ and can be simplified to the tensor form $h_r=\bm{s}\bar{G}\bm{B}$ \cite{supplement}.  

In the SO corrected basis $\Psi_{1}^{so}{-}\Psi_{4}^{so}$, we find no orbital magnetic moment contribution to the g-factor to first order in $\alpha_{i,j}$ coefficients. Second-order contributions would be much smaller than the reported shifts $\Delta g=(6\pm 1){\times}10^{-4}$ in $g_e$ \cite{Mizuochi2002}. Thus, we omit the second order SO contributions, $\alpha_{i,j}^2\ll \eta_z^e,\eta_\bot^e,\eta_\bot^a$ \cite{supplement}. We obtain the final $\textrm{V}_{\textrm{Si}}^-$ ground state spin Hamiltonian,
\begin{align}
H_B=\left(\begin{array}{cccc}
D & \sqrt{\frac{3}{8}}h_- & -i\sqrt{\frac{3}{8}}h_+ & \frac{3}{2}h_z \\
\sqrt{\frac{3}{8}}h_+ & -D+\frac{1}{2}h_z & h_- & \sqrt{\frac{3}{8}}h_+ \\
i\sqrt{\frac{3}{8}}h_- & h_+ & -D-\frac{1}{2}h_z & -i\sqrt{\frac{3}{8}}h_- \\
\frac{3}{2}h_z & \sqrt{\frac{3}{8}}h_- & i\sqrt{\frac{3}{8}}h_+ & D
\end{array}\right)\label{3}
\end{align}
in the $\Psi_{1}^{so}{-}\Psi_{4}^{so}$ basis in terms of $h_-{=}\mu_B g_\bot B_-$, $h_+{=}\mu_B g_\bot B_+$, and $h_z{=}g_{||}\mu_B B_z$ for magnetic field $\bm{B}{=}\{B_x,B_y,B_z\}$. The $z$-axis is along the defect's c-axis with $B_{\pm}{=}B_x{\pm} i B_y$. We calculated the ZFS of $2D{=}68$MHz in good agreement with experiment \cite{Kraus_NP14,Widmann_nmat15,Carter_PRBRC15,supplement}. The relativistically corrected g-factors are $g_{||}{=}g_e+(\eta_\bot^a+\eta_\bot^e)/3$ and $g_\bot{=}g_e+\eta_z^e/3+(\eta_\bot^a+\eta_\bot^e)/6$. Because of the near $T_d$ symmetry of this defect, $\eta_z^e\approx (\eta_\bot^a+\eta_\bot^e)/2$ leads to an almost isotropic g-factor $g_{||}\approx g_\bot$. In the isotropic case, the relativistic g-factors differ from $g_e$ by $\Delta g{\approx}2\eta_z^e/3$, and $\eta_z^e$ is roughly $(9\pm 1.5)\times 10^{-4}$, consistent with experiments \cite{Mizuochi2002}.

\begin{figure}[!ht]
\centering
\includegraphics*[width=8. cm]{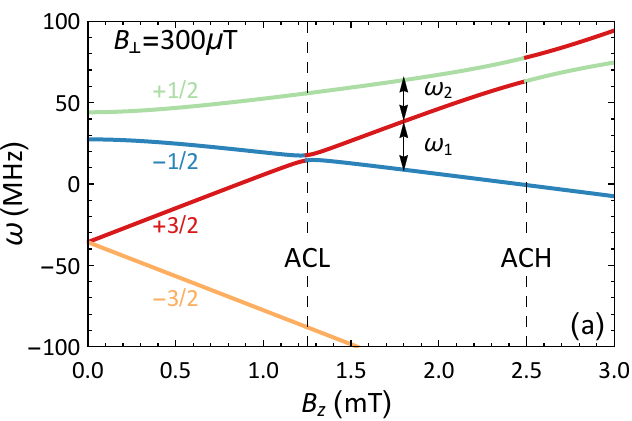} 
\mbox{\subfigure{\includegraphics*[width=3.5 cm]{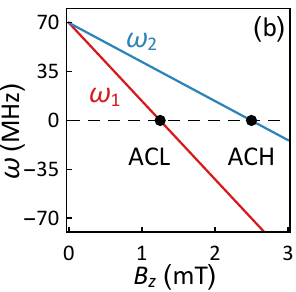}}\hspace{0.5cm}
\subfigure{\includegraphics*[width=3.5 cm]{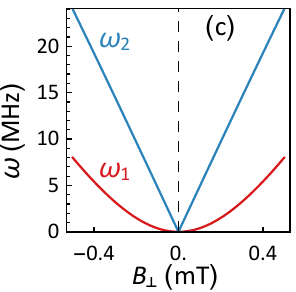}}}
\includegraphics*[width=8.5 cm]{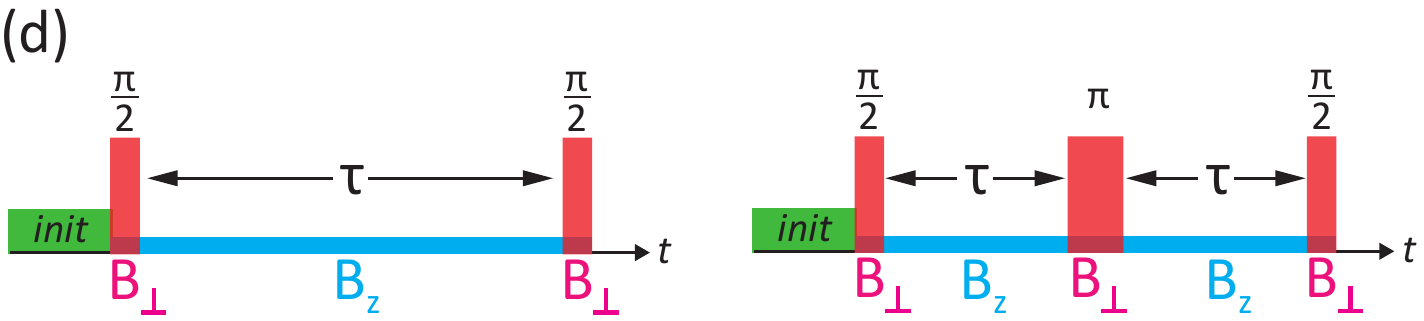}
\caption{(a) GS spin splittings versus magnetic field $B_z$ (along the c-axis) with fixed $B_\bot$ (along the basal plane). Avoided crossings ACL at the low ($\approx 1.25\textrm{mT}$) and ACH at the high ($\approx 2.5\textrm{mT}$) fields shown by vertical dashed lines. Spin projection states $\langle S_z\rangle=m_s$ are color coded. (b) GS energy splittings between spin states $m_s:3/2\leftrightarrow -1/2$ ($\omega_1$) and $m_s:3/2\leftrightarrow 1/2$ ($\omega_2$) versus $B_z$, and (c) $B_\bot$. (d) (Left) DC sensing: Ramsey pulse sequence. (Right) AC Sensing: Spin echo}\label{Fig2}
\end{figure}

The GS spin Hamiltonian in Eq.~\ref{3} can be put into the familiar single-spin ($S=3/2$) form, $H_B{=}D(S_z^2-5/4)+\mu_B\bm{S}\bar{g}\bm{B}/\hbar$, after a unitary transformation from the defect's basis to a spin-$3/2$ basis where $\bar{g}=\textrm{diag}\{g_{||},g_{||},g_\bot\}$. In the neighborhood of a level crossing where some small $B_\bot$ is present, coherent mixing/transitions that are otherwise dipole forbidden between the $m_s=3/2$ and $m_s=-1/2$ spin states can be induced without populating the auxiliary $m_s=1/2$ state. 

The resulting level repulsion between $m_s=3/2$ and $m_s=-1/2$ differing by $\Delta m_s=\pm 2$ leads to an unexpected avoided crossing when $B_\bot$ is present. It is labeled ACL in Fig.\ref{Fig2}a and occurs at a lower magnetic field than the regular avoided-crossing at higher field ACH with $\Delta m_s=\pm 1$. Near the ACL where $B_z=1.25$mT, during the Rabi oscillations between $m_s=3/2$ and $m_s=-1/2$, the $m_s=1/2$ state remains largely unpopulated due to destructive quantum interference between $3/2\leftrightarrow 1/2$ and $-1/2\leftrightarrow 1/2$. In Fig.\ref{Fig2}b, the frequencies $\omega_1$ and $\omega_2$ for transitions $3/2\leftrightarrow -1/2$ and $3/2\leftrightarrow +1/2$, respectively, both decrease linearly with a $B_z$. Level crossings corresponding to $\omega_{1,2}=0$ occur at $B_{ACL}=1.25$mT and $B_{ACH}=2.5$mT. In Fig.\ref{Fig2}c, $\omega_2$ behaves linearly with a $B_\bot$ along the basal plane as expected, whereas $\omega_1$ is almost quadratic and thus has a sharper avoided crossing in Fig.\ref{Fig2}a. This can be understood by the interference mechanism above because the second order $\Delta m_s=\pm 2$ involves two-spin resonant transitions. Note that we use a negative ZFS ($D<0$) for the GS following the recent findings \cite{Soykal_PRB2016}; however, our results remain unaffected on exchanging the signs of $m_s$ in the case of $D>0$.

Here we propose a Ramsey-type magnetic field sensing scheme (Fig.\ref{Fig2}d) using the ACL: (i) $\textrm{V}_{\textrm{Si}}^-$ spins are initialized to populate only the $m_s{=}\pm 3/2$ states by optical spin polarization at $B_z{=}0$, (ii) A field of $B_{z}{\approx}1.25$mT along the [111] c-axis moves the system into the ACL regime, (iii) A small field of $B_\bot{=}30\mu$T in the basal plane for the duration of a $\pi/2$ rotation transforms $m_s{=}3/2$ into a superposition state $(|3/2\rangle +|{-}1/2\rangle)/\sqrt{2}$, ($B_\bot$ at the ACL can be interpreted as an RF field with zero frequency) (iv) this state now evolves (precesses) freely around a small target (to be measured) magnetic field along the c-axis, accumulating a phase $\phi(\tau){=}\int_0^\tau B_{dc} dt$ over an interrogation time $\tau$, (v) A second $\pi/2$ pulse of $B_\bot$ converts the overall phase in the $m_s=3/2$ and $m_s=1/2$ states to $\langle S_z\rangle$ populations. The overall (phase induced) change in $m_s$ populations can be detected through the PL signal. 

\begin{figure}[!ht]
\centering
\mbox{\subfigure{\includegraphics*[width=4. cm]{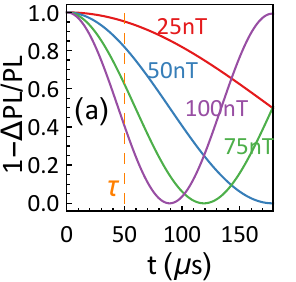}}\hspace{0.5cm}
\subfigure{\includegraphics*[width=4. cm]{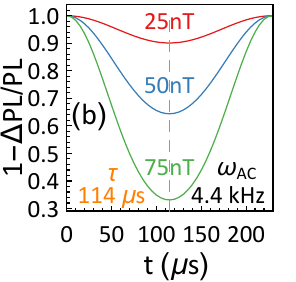}}}
\caption{(a) Change in PL versus measurement time $t$ for a range (25-100 nT) of DC fields. (b) Change in PL versus measurement time $t$ for a range (25-75 nT) of AC fields with a fixed frequency ($\omega_{\textrm{AC}}=4.4\textrm{kHz}$).
}\label{Fig3}
\end{figure}

Fig.\ref{Fig3}a shows the oscillations of the change in PL signal amplitude ($\Delta$PL) for a range of DC target fields. Smaller magnetic fields have longer oscillation periods and increasing the interrogation time $\tau$ gives an increased signal for the same small fields. Although the longer interrogation times (up to an optimal $\tau$ where $\phi(\tau)$ reaches $\pi$) improve the signal-to-noise ratio for detection of smaller fields, it ultimately will be limited by the effective $T_2^*$ (for DC) or $T_2$ (for AC) transverse relaxation times of the $\langle S_{x,y}\rangle$ components. Early measurements on single $V_{\textrm{Si}}^-$ report a lower bound of $160\mu\textrm{s}$ \cite{Widmann_nmat15} for $T_2$ times. In our evaluations we use conservative (shorter) interrogation times.

In a usual $\Delta m_s{=}\pm 1$ avoided crossing regime, the electron Zeeman energy becomes comparable to the hyperfine coupling ($\bm{S}\bar{A}\bm{I}$), and the electron spin $S$ acquire decoherence due a nearby nuclear spin $I$ by the non-secular processes, i.e., $S_x I_x$ and $S_y I_y$ \cite{Jacques_PRL09}, significantly reducing the $T_2$ times. However, the ACL here occurs between $m_s{=}3/2$ and $m_s{=}-1/2$, which differ by $\Delta m_s=\pm 2$, and the non-secular processes still involve the out-of-phase auxiliary state $m_s{=}1/2$ with larger electron Zeeman splitting. Thus they will have a much smaller probability and won't affect the interrogation times significantly because of this double resonance nature of the ACL in which at least two simultaneous nuclear spin flips are needed to change the electron spin by $\Delta m_s=\pm 2$. We note that the spin-$3/2$ defect is especially desirable for relaxation based coherent detection techniques, i.e. $T_1$-NMR \cite{Jacques_PRB13}, as a result of the reduced nuclear spin mixing effects in the ACL regime.

Next, we demonstrate AC magnetic field sensing in the ACL regime using the spin echo scheme \cite{Hahn_PR50} in Fig.\ref{Fig2}d. Fig.\ref{Fig3}b shows signals from several AC magnetic fields all with the same frequency. This frequency was chosen to achieve reasonable echo times ($2\tau$) smaller than $T_2$. The AC magnetic sensitivity is given by $\zeta_B{=}\sigma_0/(\sqrt{N}dS/dB)$ where $S{\propto}\cos^2[2\phi(\tau)]$ is the defect specific signal, $\sigma_0$ is the standard deviation per measurement and $N=T/\tau$ is the number of measurements in a one second averaging time $T$ \cite{Thomas_PRX15}. The maximum contrast between the $m_s{=}\pm 3/2$ and $m_s{=}\pm 1/2$ states is taken to be about one percent of the total average PL photon count of $40$Kcps from the defect with a solid immersion lens \cite{Widmann_nmat15}. The magnetic field response $dS/dB$ is constructed from the spin echo AC field data for $\tau=114\mu\textrm{s}$ \cite{supplement}. This gives a shot-noise limited magnetic sensitivity of $\zeta_B=40 \textrm{nT}\,\textrm{Hz}^{-1/2}$ for an AC field with frequency $\omega_\textrm{AC}=4.4 \textrm{kHz}$. We note that decreasing AC frequency increases the overall sensitivity, but will be limited by the T$_2$ relaxation time. However, a Carr-Purcell-Meiboom-Gill (CPMG) pulse-echo sequence can theoretically boost these coherence times up to the $T_1\approx 340-500\mu\textrm{s}$ \cite{Widmann_nmat15,Simin_arxiv16} relaxation times of the $\langle S_z\rangle$ components, thus permitting longer interrogation times. Therefore, sensitivities of less than $\textrm{nT}\,\textrm{Hz}^{-1/2}$ should be achievable with single $\textrm{V}_{\textrm{Si}}^-$ defect centers after optimizations involving isotopic purification and implementation optical wave-guiding to increase the photon collection efficiency. To obtain better spectral resolution, one could use a spin-locking scheme \cite{Loretz_PRL13}. 

\begin{figure}[!htp]
\centering
\mbox{\subfigure{\includegraphics*[width=3.1 cm]{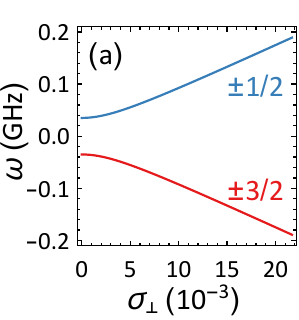}}\hspace{0.5cm}
\subfigure{\includegraphics*[width=4.9 cm]{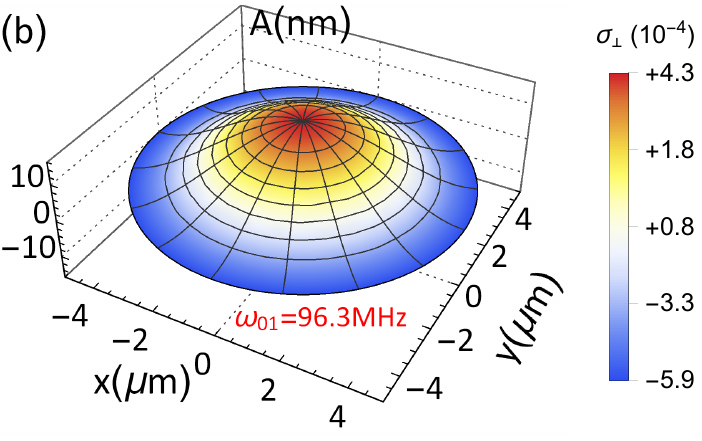}}}
\caption{(a) ZFS of the GS versus the non-axial strain $\sigma_\bot$ ($=\sigma_{xx}=\sigma_{xy}/2$ and $z{\parallel}$c-axis). (b) Fundamental mode of the SiC membrane with frequency $\omega=96.3\textrm{MHz}$ for an amplitude $|A|\approx 15 \textrm{nm}$. Surface strain shown by color. Maximum flexural strain $\sigma_m=4.34\times 10^{-4}$ corresponds to the defect placed on the surface at the center of a SiC MEMS membrane with diameter $d=10\mu m$ and thickness $h=0.3\mu m$.}\label{Fig4}
\end{figure}

We now consider the strain sensing by using a hybrid quantum system consisting of a single $\textrm{V}_\textrm{Si}^-$ defect and a SiC mechanical resonator. First, we obtain the strain Hamiltonian of the ground state up to the second order in SO coupling coefficients \cite{supplement}:
\begin{align}
H_\sigma{=}\left(\begin{array}{cccc}
D{+}\xi_1^r\Lambda^r & \xi_3\Lambda_{xy} & i\xi_{3}\Lambda^{*}_{xy} & 0 \\
\xi^*_3\Lambda^*_{xy} & -D{+}\Lambda^r \xi_2^r & 0 & -\xi^*_3\Lambda^*_{xy} \\
-i\xi^*_{3}\Lambda_{xy} & 0 & -D{+}\Lambda^r \xi_2^r & -i\xi^*_3\Lambda_{xy} \\
0 & -\xi_3\Lambda_{xy} & i\xi_3\Lambda^*_{xy} & D{+}\Lambda^r \xi_1^r
\end{array}\right).\label{4}
\end{align}
The off-diagonal terms involving $\Lambda_{xy}{=}\sigma_{xx}^E-\sigma_{yy}^E+ 2i\sigma_{xy}^E$ are obtained using the irreducible matrix elements of the strain tensor components $\sigma_{ij}^p=\langle p||\sigma_{ij}||p\rangle$ in the $C_{3\nu}$ double group \cite{Altmann2011}. In the diagonal terms, $r$ indicates a summation over the allowed MO representations, $A_1$ and $E$. This gives $\Lambda^{A_1} \xi_1^{A_1}+\Lambda^E \xi_1^E$ and $\Lambda^{A_1} \xi_2^{A_1}+\Lambda^E \xi_2^E$ in terms of $\Lambda^{A_1}{=}\sigma_{xx}^{A_1}+\sigma_{yy}^{A_1}$ and $\Lambda^{E}{=}\sigma_{xx}^E+\sigma_{yy}^E+2\sigma_{zz}^E$. Without a magnetic field, the Kramer's degeneracy of the $m_s{=}\pm 3/2$ and $m_s{=}\pm 1/2$ states under strain is conserved consistent with our expectations (see Fig.\ref{Fig4}a). The diagonal strain coupling shifts the energies of both spin multiplicities equally and therefore does not affect the ZFS. The deformation potential constants are included in the strain coupling coefficients $\xi_i$ \cite{supplement}.

For the $\textrm{V}_{\textrm{Si}}^-$ defect coupled to realistic mechanical resonators, we calculated the strain sensitivity by using typical device parameters. The defect is taken to be near the surface and at the center of a SiC membrane to maximize the strain coupling. Such devices and accurate defect placement has been already demonstrated by using various masked irradiation and smart-cut techniques \cite{Roper_Vac13}. In Fig.\ref{Fig4}a, we show the change in ZFS ($\Delta D$) with in-plane (flexural) strain $\sigma_\bot$. To be conservative, we use the smallest reported deformation potential $\Xi=11.6$ eV for bulk 4H-SiC \cite{Itoh_JAP2001}. The GS spin-strain coupling is calculated using the fundamental mode of the membrane \cite{supplement} shown in Fig.~\ref{Fig4}b. The surface flexural strain field for the fundamental mode leads to a local maximum strain of $\sigma_m{=}4.34\times 10^{-4}$ at the defect location (Fig.\ref{Fig4}b) and it results in a $\Delta D=6.87$MHz increase of the GS ZFS (Fig.\ref{Fig4}b) in the presence of a bias strain $\sigma_0$. Optically detected magnetic resonance (ODMR) can detect these variations in ZFS due to membrane oscillations. With a bias strain of $\sigma_0=10^{-2}$, we calculate the strain sensitivity to be $\zeta_\sigma{=}h_n(\omega)\omega^{-1/2}\approx 6.7\times 10^{-8}\,\, \textrm{strain}\,\textrm{Hz}^{-1/2}$. The $h_n(\omega)$ noise amplitude is estimated using the reported $2.8$ MHz (at $-10$dBm) experimental ODMR linewidth \cite{Carter_PRBRC15}. 

The advantages of the $\textrm{V}_\textrm{Si}^-$ for strain detection are: i) Roughly two orders of magnitude improvement in sensitivity over spin-1 defects \cite{Ovartchaiyapong_Nature14} due to the near $T_d$ local symmetry and Kramer's degeneracy. ii) Simpler spin-resonance detection techniques (i.e. ODMR, EPR, etc.) that does not require dynamical decoupling. This makes the $\textrm{V}_\textrm{Si}^-$ defect technologically appealing for hybrid quantum systems in realistic applications, e.g. navigation, gravimetry, and autonomous detection systems.

For the temperature dependence of the GS, we obtain a simple analytical expression \cite{supplement} leading to $dD/dT{=}1\textrm{kHz}/\textrm{K}$ change in ZFS around $T{=}300$K, in remarkable agreement with recent experiments \cite{Dyakanov2014}. This corresponds to a fractional thermal response of $dD/(DdT)=-1.4\times 10^{-5}\textrm{K}^{-1}$ and it is an order of magnitude higher than that for NV-centers in diamond \cite{Acosta2010} due to the near $T_d$ symmetry. This provides a unique opportunity for nano bio-chemical sensing with the combined benefit of increased optical penetration capabilities \cite{Smith2009} due to the $\textrm{V}_\textrm{Si}^-$ zero-phonon line that lies in the near-infrared window of biological tissue. Techniques such as the optically detected double microwave resonance between the $m_s=3/2$ and $m_s=\pm 1/2$ with an N-pulse CPMG method \cite{Wang2015} can be easily utilized to achieve this.

A fully relativistic treatment of the electronic properties of the $\textrm{V}_\textrm{Si}^-$ Si deep center defect in 4H-SiC has been used to develop opportunities for quantum metrology in this system. It has been shown that the novel features of half spin multiplet, i.e. class spin-$3/2$ quartet, defects allow for novel sensing schemes and easy-to-implement detection protocols with unique advantages that make possible sensitivities well beyond those of current technologies. Other point defects, i.e. 3d transition metal or rare-earth impurities in semiconductors, may also provide similar opportunities in quantum sensing due to their high half-spin ($S\geq 3/2$) configurations.

\begin{acknowledgments}
\"{O}.O.S. acknowledges the National Academy of Sciences Research Associateship Program. This work was supported in part by ONR and by the Office of Secretary of Defense, Quantum Science and Engineering Program.
\end{acknowledgments}


\end{document}


\setlength{\belowdisplayskip}{3pt} \setlength{\belowdisplayshortskip}{3pt}
\setlength{\abovedisplayskip}{3pt} \setlength{\abovedisplayshortskip}{3pt}

\title{Supplementary: Quantum metrology with a single spin-$3/2$ defect in silicon carbide}
\author{\"O. O. Soykal}
\email[Email: ]{oneysoykal@gmail.com}
\author{T. L. Reinecke}
\affiliation{Naval Research Laboratory, Washington, DC 20375}
\maketitle

\section{Electronic Structure}
\subsection{Electronic Hamiltonian of the deep center} 
Under the adiabatic approximation, the electronic Hamiltonian of the $\textrm{V}_{\textrm{Si}}^-$ is given as
\begin{align}
H=&\sum_{i} V_c(\bm{r}_i,\bm{R}_0){+}V_{so}(\bm{r}_i,\bm{R}_0){+}V_{hf}(\bm{r}_i,\bm{R}_0)\nonumber\\
&+\sum_{i>j}V_{ee}(\bm{r}_i,\bm{r}_j)+V_{ss}(\bm{r}_i,\bm{r}_j),\label{s1} 
\end{align}
where $\bm{r}_i$ denotes the coordinate of the i$^\textrm{th}$ electron and $\bm{R}_0$ corresponds to the equilibrium position of nuclei. The kinetic energy of the $i^\textrm{th}$ electron and the effective Coulomb potential of the interaction of the nuclei and lattice electrons with the defect electrons are summed up in $V_c$. $V_{ee}$ and $V_{ss}$ are the Coulomb repulsion potential and the spin-spin (SS) coupling of the defect electrons, respectively. $V_{so}$ is the spin-orbit (SO) coupling, and $V_{hf}$ is the hyperfine interaction between electrons and crystal nuclei.

\subsection{Spin-orbit perturbed wave functions} 
The ground state wave functions $\Psi_i^{0}$ are perturbed using the energetically close doublet states due to the spin-orbit coupling:
\begin{align}
\Psi_1^{so}(m_s=\pm 3/2)&=\Psi_1^0+\alpha_{1,8}\Psi_8+\alpha_{1,20}\Psi_{20}\nonumber\\
\Psi_2^{so}(m_s=+1/2)&=\Psi_2^0+\alpha_{2,6}\Psi_6+\alpha_{2,15}\Psi_{15}+\alpha_{2,18}\Psi_{18}\nonumber\\
\Psi_3^{so}(m_s=-1/2)&=\Psi_3^0+\alpha_{3,5}\Psi_5+\alpha_{3,16}\Psi_{16}+\alpha_{3,17}\Psi_{17}\nonumber\\
\Psi_4^{so}(m_s=\pm 3/2)&=\Psi_4^0+\alpha_{4,7}\Psi_7+\alpha_{4,19}\Psi_{19}\label{s2}
\end{align}
in terms of the coefficients given in Table \ref{Table1}. For each $\Psi_i$, we use the multi-particle orbital and spin wave functions given in the form of $\{e_x,e_y,a\}$ molecular orbital (MO) Slater determinants \cite{Soykal_PRB2016}.We calculate the first order SO perturbation coefficients by using the longitudinal and orthogonal SO coupling parameters $\lambda_{\bot,||}\cong 7$ meV \cite{Humphreys_SSC81} and \textit{ab-initio} electronic state energies $\Delta E_{1,8}\cong \Delta E_{2,6}\cong\Delta E_{3,5}\cong\Delta E_{4,7}\cong 1.032$ eV, $\Delta E_{2,15}\cong\Delta E_{3,16}\cong 1.1$ eV and $\Delta E_{1,20}\cong E_{2,18}\cong\Delta E_{4,19}\cong\Delta E_{3,17}\cong 1.153$ eV \cite{Soykal_PRB2016}.
\renewcommand{\arraystretch}{1.2}
\begin{table}[!htp]
\centering
  \begin{tabular}{lcr}
	 Coefficient & Form & Value\\
	\hline\hline
		$\alpha_{1,8}\simeq\alpha_{4,7}$ & $-\lambda_\bot/(\sqrt{2}\Delta E_{1,8})$ & $-0.0048$ \\
		$\alpha_{1,20}\simeq\alpha_{4,19}$ & $i\lambda_\bot/(\sqrt{2}\Delta E_{1,20})$ & $i 0.0043$\\
		$\alpha_{2,6}\simeq\alpha_{3,5}$ & $i\lambda_\bot/(\sqrt{6}\Delta E_{2,6})$ & $-0.0028$\\
		$\alpha_{2,15}\simeq\alpha_{3,16}$ & $-i\sqrt{2/3}\lambda_{||}/\Delta E_{2,15}$ & $-i 0.0052$\\
		$\alpha_{2,18}\simeq-\alpha_{3,17}$ & $i\lambda_\bot/(\sqrt{6}\Delta E_{2,18})$ & $i 0.0025$\\
		$\Delta E_{i,j}$ & $|E_i-E_j|$ \\
  \end{tabular}\caption{SO first order perturbation coefficients in terms of coupling parameters $\lambda_{||,\bot}$ and state energies $E_i$.}\label{Table1}
\end{table}

\section{Ground state magnetic coupling}
\subsection{Relativistic corrections to the g-tensor} 
Any change in the choice of origin is a gauge transformation leaving the g-tensor invariant, so we choose a gauge where $\bm{d}_i=\bm{r}_i$, and the redundant $j$ in the interaction Hamiltonian is dropped for each nucleus. The relativistic correction term to the g-tensor is $G_{\alpha\beta}=s_\alpha((\nabla_\gamma V) r^\gamma\delta_{\alpha\beta}-r_\alpha\nabla_\beta V)B_\beta$ where $\gamma$ is in Einstein notation and $r^\gamma{=}\{x,y,z\}$, $\nabla_\gamma{=}\partial/\partial r^\gamma$. Choosing convenient coordinate axes where nuclear potential coordinate and electron spin quantization axes are superposed, we arrive the diagonal form of $G{=}\textrm{diag}\{G_{xx},G_{yy},G_{zz}\}$ tensor with components $G_{xx}{=}y(\partial V{/}\partial y)+z(\partial V{/}\partial z)$ and its cyclic permutations. Using Wigner-Eckardt theorem \cite{Tinkham2003}, we obtain each component of the G tensor as an orbital operator diagonal in the $\{e_x,e_y,a\}$ MO basis: 
\begin{align}
G_{xx}&=\left(\begin{array}{ccc}
\eta_{z}^{e}+\eta_{\bot}^{e} & 0 & 0 \\
0 & \eta_{z}^{e} & 0 \\
0 & 0 & \eta_{\bot}^{a} \end{array}\right),\label{s3}\\
G_{yy}&=\left(\begin{array}{ccc}
\eta_{z}^{e} & 0 & 0 \\
0 & \eta_{\bot}^{e}+\eta_{z}^{e} & 0 \\
0 & 0 & \eta_{\bot}^{a} \end{array}\right),\label{s4}\\
G_{zz}&=\left(\begin{array}{ccc}
\eta_{\bot}^{e} & 0 & 0 \\
0 & \eta_{\bot}^{e} & 0 \\
0 & 0 & 2\eta_{\bot}^{a} \end{array}\right),\label{s5}
\end{align}
in terms of the non-vanishing reduced matrix elements of the nuclear potentials given in Table \ref{Table2}. Using these, the relativistic correction term becomes $\sum_i s_{x,i} G_{xx} B_x+s_{y,i} G_{yy}B_y+s_{z,i}G_{zz}B_z$ where the summation is over the defect's active electrons and $s$ is the spin-1/2 operator. 
\renewcommand{\arraystretch}{1.2}
\begin{table}[!htp]
\centering
  \begin{tabular}{cc}
	 Coefficient & Form \\
	\hline\hline
		$\eta_z^e$ & $\langle e||z(\partial V/\partial z)||e\rangle/2$ \\
		$\eta_\bot^e$ & $\langle e||x(\partial V/\partial x)||e\rangle/2{=}\langle e||y(\partial V/\partial y)||e\rangle/2$\\
		$\eta_\bot^a$ & $\langle a||x(\partial V/\partial x)||a\rangle/2{=}\langle a||y(\partial V/\partial y)||a\rangle/2$\\
  \end{tabular}\caption{Reduced matrix elements of the nuclear potentials}\label{Table2}
\end{table}

\subsection{AC magnetic field sensitivity}
AC sensitivity is defined as the minimum detectable magnetic field for $t_A{=}1\textrm{s}$ averaging time. The normalized change in PL signal ($1-\Delta \textrm{PL}/\textrm{PL}$) with respect to the magnetic field is represented by $dS/dB$ in the manuscript and it depends on the physical spin properties of the defect and of the sensing regime as well as the experimental parameters including the contrast, average photon count per second, and read-out time. 

\begin{figure}[!ht]
\centering
\includegraphics*[width=5. cm]{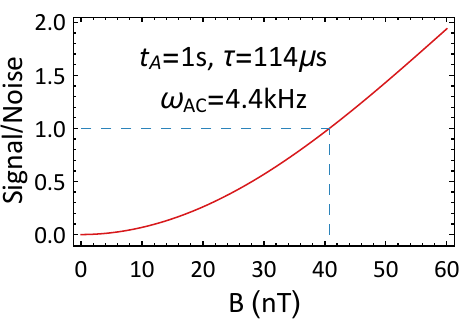}
\caption{Signal-to-noise ratio versus AC magnetic field strengths with fixed frequency $\omega_\textrm{AC}=4.4\textrm{kHz}$ for per measurement time of $\tau=114\mu\textrm{s}$. Signal is averaged for $t_A=1\textrm{s}$ for all fields.
}\label{sfig1}
\end{figure}

We examine this response function $dS/dB$ using the spin Hamiltonian in the manuscript. To do this, we calculated the normalized signal for various magnetic fields using a time dependent density matrix approach applied to a spin polarized $m_s=\pm 3/2$ state after the initialization phase. The normalized signal is then converted to an actual photon count using the previously reported $40\textrm{Kcps}$ and the $1\%{-}2\%$ contrast from this defect \cite{Widmann_nmat15}. Shot noise per measurement during $\tau=114\mu\textrm{s}$ is $\sigma_0/\sqrt{N}=200/N$ where $\sigma_0=\sqrt{4\times 10^4/N}$ is the standard deviation per measurement and $N=t_A/\tau$ is the maximum allowed number of measurements in 1s. By definition, at the minimum magnetic field $\zeta_B$, the signal is taken equal to the noise leading to the generic formula $(dS/dB)\zeta_B=\sigma_0/\sqrt{N}$ given in the manuscript. In Fig.~\ref{sfig1} we show these signal-to-noise ratios for various magnetic field strengths within a total averaging time of $t_A{=}1\textrm{s}$. Minimum magnetic field $\zeta_B=40nT$ corresponds to the signal-to-noise ratio of 1.
\section{Ground State Strain Coupling}
\subsection{Strain Hamiltonian}
Similar to the approach followed in the spin Hamiltonian, we symmetry analyze each component of the strain tensor via group theory and obtain their orbital matrix forms (Table \ref{Table3}) in the $\{e_x,e_y,a\}$ basis.
\setlength{\tabcolsep}{15pt}
\renewcommand{\arraystretch}{1.}
\begin{table}[!htp]
\centering
  \begin{tabular}{cc}
	 Strain  & Orbital Form \\
	\hline\hline
		$\sigma_{xx}$ & $\frac{1}{2}\left[\begin{array}{ccc} 0 & 0 & 0 \\ 0 & \langle e||\sigma_{xx} ||e\rangle & 0 \\ 0 & 0 & \langle a||\sigma_{xx} ||a\rangle \end{array}\right]$\\
		$\sigma_{yy}$ & $\frac{1}{2}\left[\begin{array}{ccc} \langle e||\sigma_{yy} ||e\rangle & 0 & 0 \\ 0 & 0 & 0 \\ 0 & 0 & \langle a||\sigma_{yy} ||a\rangle \end{array}\right]$\\
		$\sigma_{zz}$ & $\frac{1}{2}\left[\begin{array}{ccc} \langle e||\sigma_{zz} ||e\rangle & 0 & 0 \\ 0 & \langle e||\sigma_{zz} ||e\rangle & 0 \\ 0 & 0 & 0 \end{array}\right]$\\
		$\sigma_{xy}$ & $-\frac{1}{4}\left[\begin{array}{ccc} 0 & \langle e||\sigma_{xy} ||e\rangle & 0 \\ \langle e||\sigma_{xy} ||e\rangle & 0 & 0\\ 0 & 0 & 0 \end{array}\right]$\\
		$\sigma_{xz}$ & $-\frac{1}{4}\left[\begin{array}{ccc} 0 & 0 & \langle a||\sigma_{xz} ||e\rangle \\ 0 & 0 & 0\\ \langle a||\sigma_{xz} ||e\rangle & 0 & 0 \end{array}\right]$\\
		$\sigma_{yz}$ & $-\frac{1}{4}\left[\begin{array}{ccc} 0 & 0 & 0 \\ 0 & 0 & \langle a||\sigma_{yz} ||e\rangle \\ 0 & \langle a||\sigma_{yz} ||e\rangle & 0 \end{array}\right]$
  \end{tabular}\caption{Orbital Matrix forms of the strain components in the $\{ e_x,e_y,a \}$ basis}\label{Table3}
\end{table}

Using these in conjunction with the SO corrected wave functions of the GS in Eq.~(\ref{s2}), we obtain the general spin-strain coupling Hamiltonian of this defect:
\begin{widetext}
\begin{align}
H_\sigma{=}\left(\begin{array}{cccc}
D{+}\xi_1^r\Lambda^r & \xi_3\Lambda_{xy}+\xi_4^*\Lambda_{xz}^* & i(\xi_{3}\Lambda^{*}_{xy}+\xi_{4}^*\Lambda_{xz}) & 0 \\
\xi^*_3\Lambda^*_{xy}+\xi_4\Lambda_{xz} & -D{+}\Lambda^r \xi_2^r & 0 & -\xi^*_3\Lambda^*_{xy}+\xi_4\Lambda_{xz} \\
-i(\xi^*_{3}\Lambda_{xy}+\xi_{4}\Lambda_{xz}^*) & 0 & -D{+}\Lambda^r \xi_2^r & -i(\xi^*_3\Lambda_{xy}-\xi_4\Lambda_{xz}^*) \\
0 & -\xi_3\Lambda_{xy}+\xi_4^*\Lambda_{xz}^* & i(\xi_3\Lambda^*_{xy}-\xi_4^*\Lambda_{xz}) & D{+}\Lambda^r \xi_1^r
\end{array}\right).\label{s6}
\end{align}
\end{widetext}
with the following strain coupling coefficients:
\begin{align}
&\xi_1^{A_1}=\Xi(1+4|\alpha_{1,20}|^2)/C_1, \nonumber \\
&\xi_1^{E}=\Xi(1+3|\alpha_{1,8}|^2+|\alpha_{1,20}|^2)/C_1, \nonumber\\
&\xi_2^{A_1}=\Xi(3+2|\alpha_{2,15}|^2+4|\alpha_{2,18}|^2)/C_2,\nonumber\\ 
&\xi_2^{E}=\Xi(3+3|\alpha_{2,6}|^2+2|\alpha_{2,15}|^2+|\alpha_{2,18}|^2)/C_2, \nonumber\\ 
&\xi_3=\Xi(i\alpha_{1,8}^*\alpha_{2,6}-i\alpha_{1,20}^*\alpha_{2,18})/C_3, \nonumber\\
&\xi_4=\Xi(i\alpha_{2,15}^*\alpha_{1,8}+i\alpha_{2,15}^*\alpha_{1,20})/C_3.\label{s7}
\end{align}
The normalization constants are given by $C_1{=}2+4|\alpha_{1,8}|^2+4|\alpha_{1,20}|^2$, $C_2{=}6+4|\alpha_{2,6}|^2+4|\alpha_{2,15}|^2+4|\alpha_{2,18}|^2$ and $C_3{=}\sqrt{2 C_1 C_2}$. The deformation potential is represented by $\Xi$. Strain components $\Lambda$ are defined in the manuscript with the exception of $\Lambda_{xz}=\sigma_{xz}^E+i\sigma_{yz}^E$ that is omitted for brevity since it vanishes for the SiC membrane considered here. 

\subsection{Mechanical modes of the membrane}
We obtain each mode and its strain components of a SiC circular membrane from the solutions of the Kirchoff-Love plate theory \cite{Graff1975} along the z-axis (c-axis) given by 
\begin{align}
w=\sum_{n,m}&|A_{nm}|\left\{J_n(\lambda_{nm}r/R)-\left[J_n(\lambda_{nm})/I_n(\lambda_{nm})\right]\right.\nonumber \\
&\left.\times I_n(\lambda_{nm}r/R)\right\}\cos (\omega_{nm}t) \label{s8}
\end{align}
for small oscillation amplitudes $|A|\ll h$. For each value of $n$, $\lambda_{nm}$ is the $m^\textrm{th}$ root of the equation $I_n(\lambda_{nm})J_n^{'}(\lambda_{nm})-J_n(\lambda_{nm})I_n^{'}(\lambda_{nm})$ obtained from the clamped edge boundary condition. For the fundamental mode ($\omega_{01}$), it is $\lambda_{01}=3.196$. $J_n$ and $I_n$ are the $n^\textrm{th}$ order regular and modified Bessel functions of the first kind. The only non-vanishing strain tensor components are $\sigma_{xx}=-z\partial^2 w/\partial x^2$, $\sigma_{yy}=-z\partial^2 w/\partial y^2$, and $\sigma_{xy}=-2z\partial^2 w/\partial x \partial y$. The frequency of each mode is given by $\omega_{nm}^2=(2\lambda_{nm}/d)^4 E/(\rho h)$. The bending stiffness $E=Y h^3/\left(12(1-\nu^2)\right)$ is obtained in terms of Young modulus $Y=748$GPa, Poisson ratio $\nu=0.45$, and material density $\rho=3211\textrm{kg}/\textrm{m}^\textrm{3}$ \cite{Gmelins1959} of 4H-SiC. 

\begin{figure}[!hb]
\centering
\mbox{\subfigure{\includegraphics*[width=4. cm]{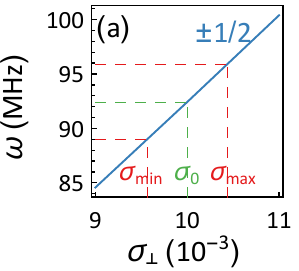}}\hspace{0.5cm}
\subfigure{\includegraphics*[width=4. cm]{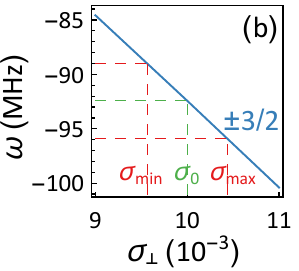}}}
\caption{The optically detectable frequency change due to very small variations in membrane strain is shown for Kramer's degenerate (a) $\pm 1/2$ and (b) $\pm 3/2$ spin states.
}\label{sfig2}
\end{figure}

To achieve maximum sensitivity, we considered an initially strained SiC mechanical resonator by a bias strain of $\sigma_0=10^{-2}$. Various external effects such as acceleration/deceleration of the device, gravitation, or added mass of a molecule can create changes on the membrane strain and can be optically detected by measuring the ZFS with ODMR. For a membrane only displaced $15$nm by any of these environmental forces and thus attained a small increase in strain to $\sigma_m$, the change in energy of $\pm 1/2$ and $\pm 3/2$ spin states are shown in Fig.\ref{sfig2} leading to an increase of the splitting between them. 

\section{Temperature dependence of ZFS}
Thermal expansion of the SiC crystal changes the distances and bonding angles between the neighboring carbon atoms around the defect. The ground state ZFS is determined mostly by the dipolar spin-spin interactions between the defect's active electrons \cite{Soykal_PRB2016}. The dipolar coupling results from each spin generating a magnetic field that is oriented parallel to the electron spin vector. Two spins located around the vacancy experience each other's magnetic field that depends on the orientation of both magnetic dipoles. The dipole-dipole coupling strength depends strongly on the spin-spin distance $1/r^3$ and on the dipolar angle $\theta$ between the axis connecting the two spins and the c-axis of the defect. Therefore, the thermal expansion of the lattice around the defect leads to a change in the ZFS of the GS determined by the dipolar spin-spin interactions \cite{Soykal_PRB2016}, and as a result the $\textrm{V}_\textrm{Si}^-$ defect can be used as a temperature sensor. We note that in general one should consider the thermally induced strain effects on the ZFS; however for defects close to the surface and away from the substrate interface the host-material can expand nearly freely and this effect can be neglected.
 
The axial thermal expansion coefficients (in Celsius) of 4H-SiC are reported \cite{Li_JAP1986} as $c_{11}{=}3.21\times 10^{-6}+3.56\times 10^{-9}T$ for the a-axis and $c_{33}{=}3.09\times 10^{-6}+2.63\times 10^{-9}T$ for the c-axis of the defect. We calculate the defect's lattice expansion with respect to temperature and its effect on the neighboring carbon atom positions by $x_{i,k}(\Delta T)=x_{i,k}(T_0)(1+c_{kk}(T_0)\Delta T)$ with respect to the center of the defect. $x_{i,k}$ represents the $i^{\textrm{th}}$ carbon atom's position on the $k^{\textrm{th}}$ principal axis ($a$- or $c$-axes). We start from zero temperature using the \textit{ab initio} bond lengths \cite{Soykal_PRB2016} $a=2.0547$\AA , $d=2.0577$\AA, and bond angles $\theta_{ad}=109.423^\circ$, $\theta_{ab}=109.522^\circ$ for a relaxed structure. The temperature dependent ZFS equation is obtained as 
\begin{align}
D=&\gamma_0\left[\eta_{ad}\langle r_{ad}(\Delta T)^{-3}\rangle(1-3\cos^2\theta_{ad}(\Delta T))\right.\nonumber \\
&\left. +\eta_{ab}\langle r_{ab}(\Delta T)^{-3}\rangle\right]/4
\end{align} 
in terms of parameters $\gamma_0=\mu_0g^2\mu_B^2/(4\pi)$, $\eta_{ab}=1.443$ and $\eta_{ad}=1.557$ \cite{Soykal_PRB2016}. The distance between the two carbon atoms along the c-axis and on the basal plane is given by $r_{ad}$ whereas the distance between the two basal plane carbons is $r_{ab}$.
